\documentclass[reprint,
bibnotes,
 amsmath,amssymb,
 prl,
]{revtex4-2}

\usepackage{graphicx}
\usepackage{dcolumn}
\usepackage{bm}
\usepackage{xcolor}
\usepackage{lineno}
\setcitestyle{square, numbers,compress}

\usepackage{graphicx}

\usepackage{xr-hyper}
\usepackage{hyperref}
\makeatletter

\begin{document}

\preprint{APS/123-QED}
\raggedbottom

\title{Magnetically Tuned Metal-Insulator Transition in LaAlO$_3$/SrTiO$_3$ Nanowire Arrays}


\author{Ranjani Ramachandran$^{1, 2}$}
\author{Shashank Anand$^{3, 4}$}
\author{Kitae Eom$^5$}
\author{Kyoungjun Lee$^5$}
\author{Dengyu Yang$^{1, 2}$}
\author{Muqing Yu$^{1, 2}$}
\author{Sayanwita Biswas$^{1, 2}$}
\author{Aditi Nethwewala$^{1, 2}$}
\author{Chang-Beom Eom$^5$}
\author{Erica Carlson$^{3,4}$}
\author{Patrick Irvin$^{1, 2}$}
\author{Jeremy Levy$^{1, 2}$}
\email{jlevy@pitt.edu}

\affiliation{
1. Department of Physics and Astronomy, University of Pittsburgh, Pittsburgh, Pennsylvania 15260, USA
}%
\affiliation{2. Pittsburgh Quantum Institute, Pittsburgh, Pennsylvania 15260, USA
}%
\affiliation{3. Department of Physics and Astronomy, Purdue University, West Lafayette, IN 47907, USA}
\affiliation{4. Purdue Quantum Science and Engineering Institute, West Lafayette, IN 47907, USA}
\affiliation{5. Department of Materials Science and Engineering, University of Wisconsin–Madison, Madison, Wisconsin 53706, USA}
\date{\today}

\begin{abstract} 
A wide family of two dimensional (2D) systems, including stripe-phase superconductors, sliding Luttinger liquids, and anisotropic 2D materials, can be modeled by an array of coupled one-dimensional (1D) electron channels or nanowire arrays. Here we report experiments in arrays of conducting nanowires with gate and field tunable interwire coupling, that are programmed at the LaAlO$_3$/SrTiO$_3$ interface. We find a magnetically-tuned metal-to-insulator transition in which the transverse resistance of the nanowire array increases 
by up to four orders of magnitude. To explain this behavior, we develop a minimal model of a coupled two-wire system which agrees well with observed phenomena. These nanowire arrays can serve as a model systems to understand the origin of exotic behavior in correlated materials via analog quantum simulation.

\end{abstract}

\maketitle

\subsubsection{\label{sec:level1a}Introduction}

Electrons confined to 1D geometries can potentially exhibit strong interactions and show exotic properties like spin-charge separation  \cite{Chang2003-ql}, charge fractionalization \cite{Steinberg2007-ih} and Luttinger liquid behaviour \cite{Luttinger1963-uz}. Models based on coupled networks of 1D nanowires have generated interest as they can potentially host charge density waves \cite{Quan2024-wy}, smectic metallic phases \cite{Emery2000-qb,Jose2022-mt}, BKT transition\cite{Kundu2021-ww}, sliding Luttinger liquid phases \cite{Mukhopadhyay2001-uz}, fractional quantum hall effect  \cite{Kane2002-pi}, and more. Stripe-like correlations are often found in higher-dimensional systems  \cite{Emery1999-py}. For example, organic conductors showing anisotropic transport in different directions can be regarded as a system of weakly coupled nanowires \cite{Lerner2012-te,Georges2000-yu}. The well-known stripe instabilities in high-$T_c$ superconductors like La$_{2-x}$Sr$_x$CuO$_4$  \cite{Cheong_undated-qq}  have a rich interplay with the superconductivity in cuprates \cite{Emery1999-tn,Emery1997-cw}. Stripe-like growth of insulating nanodomains have been reported near the metal-insulator transition (MIT) in rare-earth nickelates such as NdNiO$_3$ \cite{Mattoni2016-xn} and Mott insulators like Sr$_3$(Ru$_{1-x}$Mn$_x$)$_2$O$_7$ \cite{Kim2010-fe}. The formation of pairs of charge-ordered stripes was observed on La$_{1-x}$Ca$_x$MnO$_3$, and is expected to have an impact on its magnetotransport properties \cite{Mori1998-ca}. 

The experimental realization of controllable coupled wire systems is highly limited \cite{Yu2023-sg}. The motif of analog quantum simulation can provide experimental avenues for addressing this important family of electronic materials.  Coupled with theoretical frameworks that can guide and interpret experimental findings, the use of reconfigurable quantum materials offers the possibility of new insights into these important problems. In this work, we create a system of coupled nanowires using a reconfigurable analog quantum simulation platform based on the complex-oxide heterostructure  LaAlO$_3$/SrTiO$_3$ (LAO/STO) \cite{Pai2017-rh,Levy2022-bl}.   
This interface hosts a 2D electron system (2DES) that can be locally switched from insulating to conducting state reversibly using either conductive atomic force microscope (c-AFM)
lithography  \cite{Cen2008-gt, Levy2014-nc, Bi2010-vy} or 
ultra-low voltage electron-beam lithography (ULV-EBL).  These techniques enable arbitrary patterns of conducting regions to be defined with a resolution  
$\sim 10$~nm  \cite{Yang2020-rs}. The created structures are tunable with a back-gate voltage, which enables systematic study of the interwire coupling. Once the 2DES is formed at the LAO/STO interface, it undergoes a superconducting phase transition with temperature $T_c\approx 200 - 300$ mK and a critical field of 100 - 200 mT \cite{Reyren2007-sz,Pai2018-ph}.

\begin{figure*}
\includegraphics{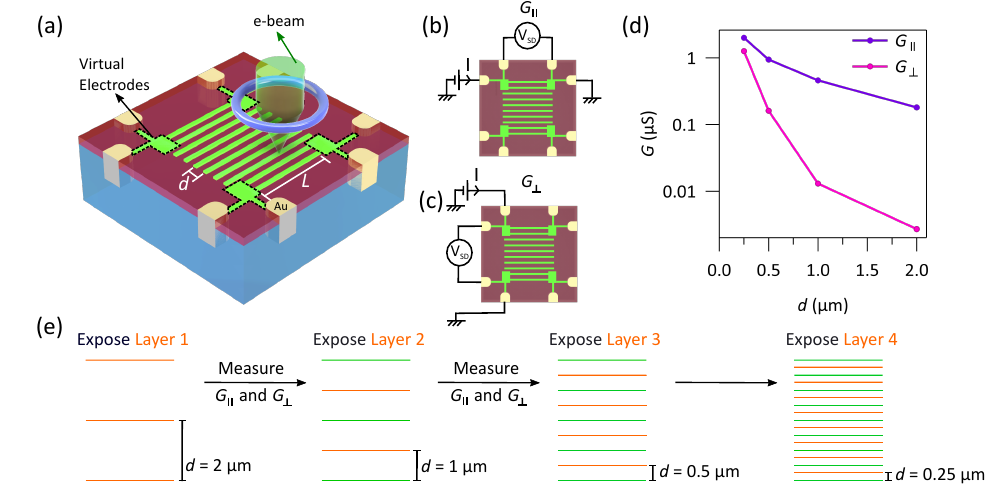}  
\caption{(a) Schematic illustration (not to scale)  of the nanowire array(green) at the LaAlO$_3$(red)/SrTiO$_3$(blue) interface. Virtual electrodes are indicated by dotted black boundary. Au electrodes are deposited at the LAO/STO interface. Configuration to measure transport properties (b) parallel and (c) perpendicular to the direction of wires. (d)Conductance {\em vs.} the interwire spacing, d, measured at room temperature for both the configurations. (e) Schematic of the in-situ measurement sequence during device exposure. Current exposure layer is shown in orange and previously exposed nanowires are shown in green.}
\label{fig:fig1}
\end{figure*}

\subsubsection{\label{sec:level1b}Experimental System}
LAO/STO heterostructures with 3.4 unit cells of LAO are grown by pulsed laser deposition on TiO$_2$-terminated STO (001) substrates, as described in Ref. \cite{Nethwewala2024-vf}. At this LAO thickness, the interface is intrinsically insulating  \cite{Ohtomo2004-ed,Thiel2006-ds} and can be made locally conductive upon exposure using ULV-EBL. Ar-ion milling and standard photolithography techniques are employed to make 4 nm Ti/ 20 nm Au interface electrodes that contact the LAO/STO interface. Ti/Au surface bond pads are made without the ion milling step and extend out from these interface electrodes. Al wirebonds are used to contact the bond pads to the chip carrier.  The sample is affixed to a chip carrier with silver paste. Experiments are performed on four devices, which show consistent behavior. We first focus on ``Device 1''. 

Fig. \ref{fig:fig1}(a) provides a schematic representation of Device 1. The
electron beam exposed
structures (indicated in green) define the main nanowire array device as well as ``virtual electrodes'' that are used to contact the nanowire array to the Au interface electrodes. The main device consists of an array of parallel 1D conducting nanowires of length $l=$ 38 $\mu$m for Device 1 and interwire spacing $d$. Several 1D nanowires come in contact with each of the four contacts located at each corner. The actual layout file used for experiment is attached in the Supplementary material.

Figs. \ref{fig:fig1}(b) and (c) show the four-terminal configuration  used to measure $G_{||}$ and $G_\perp$, the conductance parallel and perpendicular to the nanowires, respectively. The main device is created through exposure of electron beam of acceleration voltage of 100 V and a line dose of 100 pC/cm at room temperature and a pressure $P<5\times 10^{-6}$ mbar. Layer 1 consisting of lines with $d= 2~\mu$m is exposed first and $G_{||}$ and $G_\perp$ are then measured in-situ. Subsequent exposure of Layer 2 creates lines in between the pre-existing lines, which reduces the spacing to 1 $\mu$m. This exposure step is followed by measurement of $G_{||}$ and $G_\perp$. This process of sequentially halving $d$ and measuring the conductance is repeated until $G_{||}$ and $G_\perp$ are within a factor of two of one another. Fig. \ref{fig:fig1}(d) shows $G_{||}$ almost doubling as $d$ is halved because the number of nanowires touching the virtual electrodes is doubled (Fig. \ref{sebl}(c)). $G_\perp =$ 2.7 nS at $d=$ 2 $\mu$m, which is within the expected background conductance measured across the insulating sample. $G_\perp $ eventually increases to 1.25 $\mu$S when  $d=$ 250~nm, indicating that this is the spacing at which electrons can tunnel between the wires at room temperature and lead to a significant transverse conductance. The device, with $d=$ 250 nm is transferred into a dilution refrigerator. All the following experiments were performed in a dilution fridge with mixing chamber temperature of 30 mK unless otherwise specified. 

\begin{figure*}
\includegraphics[scale=1]{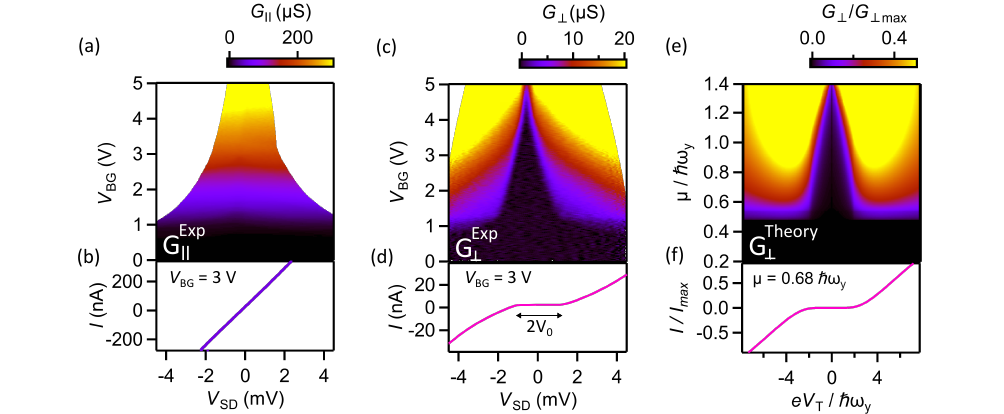}
\caption{(a) $G_{||}$ and  (c) $G_\perp$ for Device 1 as a function of $V_{SD}$ and $V_{BG}$, taken at B = 1 T (b) IV curve along the nanowires at $V_{BG} = 3V$  and (d) Corresponding IV curve transverse to the nanowires at $V_{BG} = 3 V$. (e) and (f) show $G_\perp$  and IV curve obtained from the theory model calculated for $m^*=m^*_x=m^*_y=1.9m_e$, $l_y = 17$nm and $d =$ 120 nm. }
\label{fig:fig2}
\end{figure*}

\subsubsection{\label{sec:level1c}Experimental Results}

Fig. \ref{fig:fig2} shows the conductance obtained by numerically differentiating the current-voltage (``IV'') curves taken at $B =$ 1 T perpendicular to sample plane, sufficiently large to disrupt superconductivity in  LAO/STO.  A back-gate voltage $V_{BG}$ is applied to adjust the chemical potential of the device. IV curves taken parallel to the nanowires show ohmic behavior, with $G_{||}$ increasing monotonically with $V_{BG}$ as expected (Fig. \ref{fig:fig2}(a,b)). Fig. \ref{fig:fig2}(d) shows an insulating tunnel barrier of strength $2V_0$ in the IV curve transverse to the nanowires.  When $|V_{SD}| < V_0$ ($\approx$ 1.5 mV in this curve) the transport perpendicular to the nanowires is highly insulating. When $|V_{SD}| > V_0$, there is a non-linear increase in current, suggestive of tunneling conductance. Fig. \ref{fig:fig2}(c) shows $G_\perp$ as a function of $V_{BG}$ and $V_{SD}$. When $V_{BG} < 1 V$, $G_\perp$ is zero for all values of $V_{SD}$. When $V_{BG} > 1 V$, we observe a gate-dependent tunneling barrier. $V_0$ is the threshold value of $V_{SD}$ at which interwire conduction begins and is related to the energy barrier for interwire electron hopping. As $V_{BG}$ increases, it tunes the system through three regimes: (1) an array of independent 1D nanowires ($V_{BG} < 1 V$), (2) coupled quasi-1D system ($1 V < V_{BG} < 5 V$), and (3) a nearly isotropic 2D system ($V_{BG} > 5 V$). It should be noted that back-gating is known to be hysteretic in LAO/STO. Further, $V_0$ increases with decreasing temperature from 4 K to 100 mK and $G_\perp$ shows power-law behaviour in this range (End matter Fig. \ref{d3} and \ref{powerlaw}). Similar characterisation of the device in the superconducting regime $T <$ 50 mK and $B =$ 0 T is also discussed in the End matter (Fig. \ref{fig:sc})

\begin{figure*}
\includegraphics{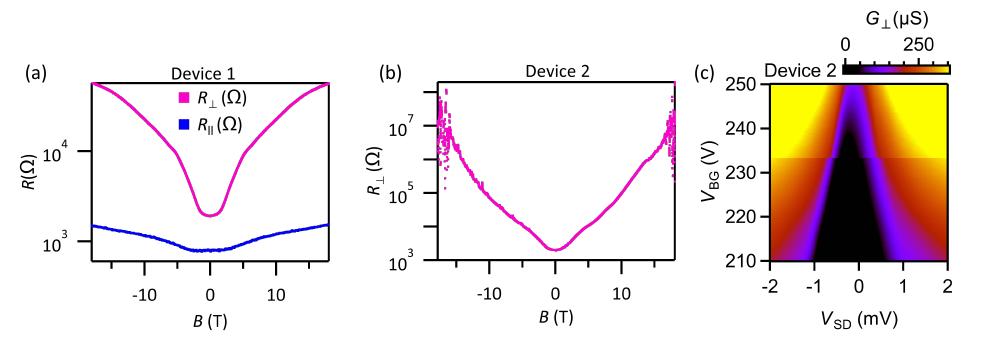}
\caption{Out-of-plane magnetic field dependence of (a) $R_{||}$ and $R_\perp$ at $V_{BG}=$ 25 V for Device 1 (b) $R_\perp$ at $V_{BG}=$ 250 V for Device 2, showing more than 10000x change in $R_\perp$ (c) $G_\perp$ for Device 2 as a function of $V_{SD}$ and $V_{BG}$, taken at B = 1 T}
\label{fig:Magnetoresistance}
\end{figure*}

Application of an out-of-plane magnetic field $B$ (Fig. \ref{fig:Magnetoresistance}(a)) causes $R_\perp$ to increase by two orders of magnitude, while $R_{||}$ changes by less than a factor of two. This large magnetoresistance in $R_\perp$ can be further enhanced (suppressed) upon lowering (increasing) $V_{BG}$(Fig. \ref{MagnetoRvsVBG}), as expected from the $V_{BG}$ dependence of $V_0$ described before. Fig. \ref{fig:Magnetoresistance}(b) is from Device 2 (same geometry as Device 1) where $V_{BG}$ was tuned to achieve four orders of change in $R_{\perp}$, beyond which the current is too small to be measured by the instrument. Fig. \ref{fig:Magnetoresistance}(c), shows $V_0$ decreasing with $V_{BG}$ in Device 2 as seen in Device 1. Such non-saturating magnetoresistance has been reported in semimetals \cite{Huang2015-mf,Xiong2015-ex,Narayanan2015-dm,Liang2015-ae,Novak2015-fa} as well as complex oxides \cite{David2015-gw,Mallik2022-zw}. Recent reports on oxide systems suggest a connection between large magnetoresistance and mobility inhomogeneity in samples, and further suggest defect engineering using oxygen vacancies as a way to engineer systems with high magnetoresistance \cite{Christensen2024-sn,Song2015-dm}. Here, the creation of nanowires act as a controllable way of creating such a mobility inhomogeneity.

\subsubsection{\label{sec:level1d}Theoretical Model}

Our theoretical modeling of the nanowire array aims to capture some distinctive transport features seen in the wires, including the sharp onset of conduction, and the transition to a highly insulating state as a function of an external electric field and an out-of-plane magnetic field.  For a single wire, we adopt a waveguide model of non-interacting electrons which was previously shown to successfully model ballistic quantized transport in single quasi-1D LAO/STO nanowire devices~ \cite{Annadi2018-vb}. We assume translational invariance along the nanowire direction ($x$), parabolic confinement along the transverse direction ($y$), and half-parabolic confinement along the vertical ($z$) direction: $V(z)=\frac{1}{2}m\omega_z^2z^2$ for $z\geq 0$ and $V(z)=\infty$ for $z<0$.
The corresponding Hamiltonian is as follows:  
\begin{align}
    \begin{split}
      \mathcal{H}_{wire}&=\frac{1}{2m^*}(p_x-eBy)^2+\frac{p_y^2}{2m^*}+\frac{p_z^2}{2m_z^*
      }  \\& \quad+\frac{1}{2}m^*\omega_y^2y^2 +\frac{1}{2}m^*_z \omega_z^2z^2  -g\mu_BBs \label{eqn:hwire}
    \end{split}    
\end{align}
where the frequency $\omega_y=\hbar/m^*_yl_y^2$ corresponds to an effective wire width $l_y$. Referring to  Ref. \cite{Annadi2018-vb}, we choose the Landé g-factor to be $g=0.62$. The eigenenergies are of  the form
\begin{align}
\begin{split}
E_n(B)&=(n_y+1/2)\hbar \Omega_B+((2n_z+1)+1/2)\hbar\omega_z  \\& \quad -
g\mu_B Bs+\frac{\hbar^2k_x^2\omega_y^2}{2m^*\Omega_B^2}~.  \label{eqn:en} 
\end{split}
\end{align}
where $\Omega_B$ is the effective frequency of the harmonic oscillator levels that arise due to both the confinement of the wire and the applied magnetic field and is given by $\Omega_B=\sqrt{\omega_y^2+(eB/m^*)^2}$. We assume that the 
electrons occupy only the lowest subband along the z direction. We therefore set $n_z=0$. The electron waveguide model agrees well in the single-electron regime with previous experiments on single wires on LAO/STO\cite{Annadi2018-vb} and is exactly solvable even in the presence of an external magnetic field.

\begin{figure*}
\includegraphics{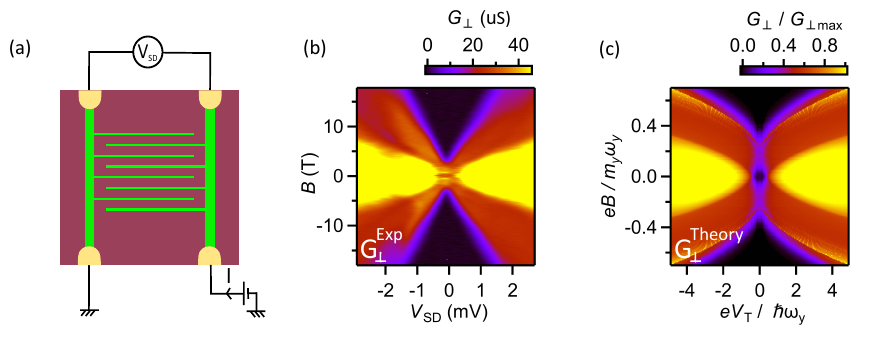}
\caption{(a) Schematic of Device 4 with interdigitated nanowire array.  (b) $G_\perp$ as a function of $V_{SD}$ and $B$ at and $V_{BG}$ = 2 V.  (c) Corresponding results from the theoretical model calculated for $m^*=m^*_x=m^*_y=1.9m_e$, $l_y = 17$ nm, $d = $120 nm, $\mu = 1.38 \hbar\omega_y$ and $E_0 =  1.58 \hbar\omega_y$ }
\label{fig:IDT}
\end{figure*}

We consider two nanowires separated by a distance $d$. We use the following interwire Hamiltonian to calculate tunneling between wires. 
\begin{align}
\begin{split}
  \mathcal{H}_{inter}=&\frac{1}{2m^*}\Big(p_x-eB(y-\frac{p_x\omega_c}{m^*\Omega_B^2}) \Big)^2 \\
  &+\frac{p_y^2}{2m^*} -eV_T\frac{(y-y_{0})}{d-y_0} 
  +E_0-g\mu_BBs
  \label{hinter}
\end{split}
\end{align}
Here, $E_0$ is the zero-field potential barrier height measured from the bottom of the wire,
$d$ is the interwire spacing and $V_{T}$ is the transverse bias applied between the wires. In the presence of a magnetic field, the electron experiences a momentum-dependent shift 
in its guiding center, which in turn influences the shape of the interwire potential 
(see [Anand et al., to be published] for details). Furthermore, we assume that the chemical potential of the wire set by the backgate voltage  is comparable to $E_0$. In this limit, the transmission barrier is more sensitive to the structure of the potential landscape and the application of external fields can alter the rate of tunneling.
The transverse bias, $V_T$ enhances tunneling by making the barrier thinner while a magnetic field increases confinement within the wires by increasing the barrier height. We assume that the tunneling process obeys energy  and momentum conservation. 
Significant tunneling of electrons is activated when the chemical potential of the first nanowire is higher than the potential at the edge of the second nanowire, $\mu\geq V_{inter}(d)$. This threshold occurs at a value of $V_T=V_0$ given by:

\begin{eqnarray}
\mu && =V_{inter}(y=d,k_x=\pm k_F,s) \nonumber\\ 
&&=\frac{1}{2}m^*\omega_c^2\Big(d-\frac{p_x\omega_c}{m^*\Omega_B^2}\Big)^2 - \hbar k_x\omega_c\Big(d-\frac{p_x\omega_c}{m^*\Omega_B^2}\Big) \nonumber \\ 
&& \quad -eV_0+E_0-\mu_B gBs
\end{eqnarray}

This expression captures many of the observed features. For example, since $V_{BG}$ is proportional to $\mu$, the linear decrease in $V_0$ with $V_{BG}$ follows from this expression. Fig. \ref{fig:fig2}(e,f) show the interwire conductance and IV curves calculated from this model. Comparing this with Fig. \ref{fig:fig2}(c,d) shows that our model qualitatively predicts the interplay between $V_{BG}$ and $V_{SD}$ well. Further details of this theoretical description are presented elsewhere [Anand et al.,](yet to be published).
The theoretical model suggests that the essential features can be captured by tunneling between two nanowires. To test this hypothesis, we created an interdigitated device (``Device 4'') with 4 pairs of nanowires connected in parallel, as shown in Fig. \ref{fig:IDT}(a). The measured conductance, $G_\perp$, in this device will consist of the cumulative signal of multiple nearest interwire hoppings. Transport experiments on this device also show  coupling between the wires that is similarly tunable with $V_{BG}$(Fig. \ref{d4}) and $B$. Fig. \ref{fig:IDT}(b) was obtained at $V_{BG} = 2$ V, chosen such that $V_0 = 0$ at  $B = 0$.  $V_0$ increases with $B$. Fig. \ref{fig:IDT}(c), calculated from the theory model captures the major features in the Fig. \ref{fig:IDT}(b). The finer magnetic field dependent features are further described in  [Anand et al., forthcoming].


\medskip
\subsubsection{\label{sec:level1e}Discussion and Conclusion}

In summary, we have investigated a system of coupled 1D nanowires with gate-tunable interwire coupling, programmed in LAO/STO heterostructures. At milli-Kelvin temperatures, the devices show highly conductive transport along the nanowires, while along the transverse direction, the interwire coupling is strongly tuned with back-gate voltage and magnetic field.  
We observe a large positive non-saturating magnetoresistance in the transverse direction under the application of out-of-plane magnetic field, which can be further enhanced by lowering the back-gate voltage. A two-wire model is introduced to describe these effects,
accounting for a conductor-to-insulator transition as a function of magnetic field. 
This LAO/STO-based analog quantum simulation platform can be employed to explore the physics of coupled quasi-1D systems, potentially impacting our understanding of High  $T_c$ superconductors, organic quasi-1D conductors, Luttinger liquids, and charge-ordered phases. Meandering stripes are hypothesized to destabilize the charge-density-wave state that usually competes with superconductivity, and lead to enhancement in superconductivity \cite{Kivelson1998-gk,Carlson2004-ye,Emery1999-py}. Our experiments indicate that superconductivity of LAO/STO is preserved during interwire tunneling(Fig.\ref{fig:sc}). Along with the ability to create meandering nanowires with arbitrary shapes, the complex oxide-based strategy of engineering coupled 1D nanowires open avenues to explore the effect of stripe phase on high-$T_c$ superconductivity, and test other theoretical predictions as well \cite{Morpurgo2024-sk,Kane2002-pi}.

\medskip
\textbf{Supporting Information} \par
Supporting Information is available as Supplementary information.pdf.

\medskip
\textbf{Acknowledgements} \par
JL, PI, C-BE, and EC acknowledge support from from the Department of Energy under grant DOE-QIS (DOE DE-SC0022277). 
EC acknowledges support from NSF Grant no. DMR-2006192.
C.B.E. acknowledges support for this research through the Gordon and Betty Moore Foundation’s EPiQS Initiative, Grant GBMF9065 and a Vannevar Bush Faculty Fellowship (ONR N00014-20-1-2844). Transport measurement at the University of Wisconsin–Madison was supported by the US Department of Energy (DOE), Office of Science, Office of Basic Energy Sciences (BES), under award number DE-FG02-06ER46327 (C.B.E.).
\medskip

\textbf{Conflict of Interest} \par
The authors declare no conflict of interest.

\newpage
\bibliography{nanowires,stokto}

\newpage
 
\subsection{\label{sec:level2}End Matter}

\subsubsection{Methods}
R vs B data in Fig.\ref{fig:Magnetoresistance} was obtained using lockin detection of in-phase/resistive component at 13 Hz. All other conductance color plots were obtained by numerically differentiating the IV curves to obtain G = dI/dV at different values of parameters as indicated. 

\subsubsection{Superconductivity in Device 1}
\begin{figure*}
\includegraphics{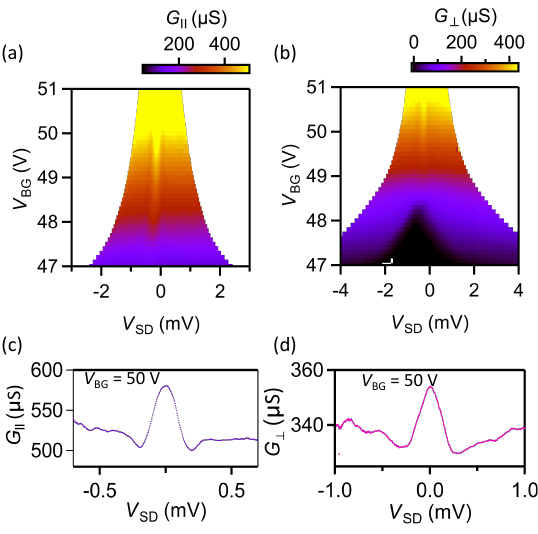}
\caption{(a) $G_{||}$ and  (c) $G_\perp$ conductance obtained from IV curves taken at B = 0 T, T $<$ 50 mK as a function of $V_{BG}$. (b) and (d) Linecuts at $V_{BG} = 50$ V taken from (a) and (c) respectively, showing enhanced conductance near zero bias due to superconducting state}
\label{fig:sc}
\end{figure*}

Fig. \ref{fig:sc}(a) and (c) show a zero bias enhancement in $G_{||}$, which is indicative of superconductivity in LAO/STO. Non-zero resistance in superconducting state for 1D systems is attributed to formation of normal state hotspots\cite{Tinkham2003-gm}, quantum phase slips\cite{Giordano1988-hi,Lau2001-fw} and has been reported before in LAO/STO nanowires\cite{Pai2018-ph,Tang2020-cz}. In \ref{fig:sc}(b) and (d), $G_\perp$ also shows a similar superconductivity once the tunneling barrier is overcome and $V_0 =$ 0, indicating the ability of phase coherent electrons to tunnel across the nanowires. This can potentially give us the ability to study the impact of stripe correlations on superconductivity.

\subsubsection{Temperature dependence of \boldmath\texorpdfstring{ $G_\perp$}{Gperp}}

\begin{figure*}
\includegraphics{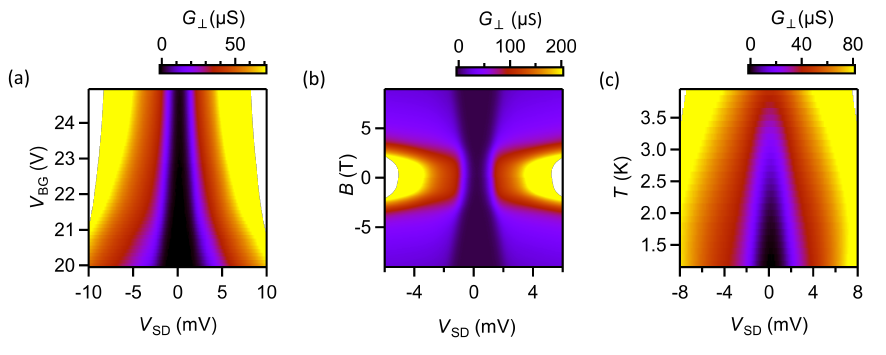}
\caption{Data from Device 3 with  $l=$ 38 $\mu$m and $d=$ 250 nm: (a) $G_\perp$ as a function of $V_{SD}$ and $V_{BG}$ taken at B = 0 T, T $=$ 1.2 K (b) $G_\perp$ as a function of $V_{SD}$ and B taken at $V_{BG} = 85$ V, T $=$ 100 mK (c) $G_\perp$as a function of $V_{SD}$ and T taken at $V_{BG} = 20$ V, B $=$ 0 T}
\label{d3}
\end{figure*}

\begin{figure*}
\includegraphics{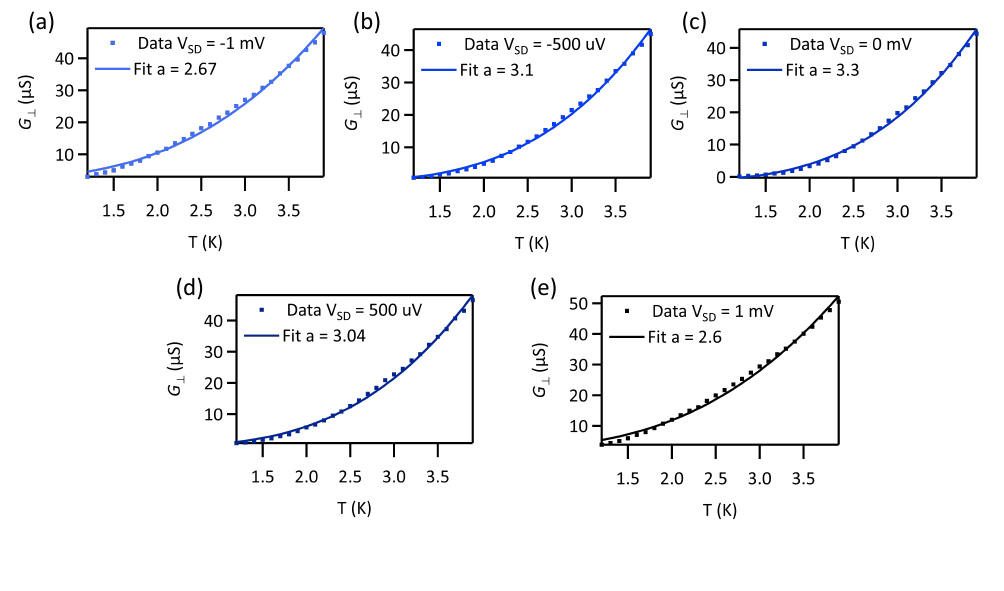}
\caption{Vertical linecut from \ref{d3}(c) and the corresponding power law fits at $V_{SD}$ (a) -1 mV (b) -500 uV (c) 0 V (d) 500 uV (e) 1 mV. Dotted line shows the experimental data and solid line shows the power law fit}
\label{powerlaw}
\end{figure*}

\ref{d3}(a) and (b) show the gate tunable tunneling barrier at much higher temperature of T $=$ 1.2 K and  magnetic field dependence of tunneling barrier respectively. \ref{d3}(c) shows the tunneling barrier increase with lowering temperature. We investigate this temperature dependence further by taking the vertical linecuts of \ref{d3}(c) and fitting it to a power law function defined as follows
\begin{align}
G_{\perp}(T)&=  K T^{a} + c
\label{eqn:fit}
\end{align}
where c is a small offset, K is an amplitude and $a$ is the power in the power law fit. Fig. \ref{powerlaw} shows the power law fit at different $V_{SD}$.Such power law dependence of conductance could potentially be indicative of luttinger liquid behavior as reported before in carbon nanotubes\cite{Bockrath1999-ld,Yao1999-lp} and more recently in twisted moire materials\cite{Wang2022-nh,Yu2023-sg}. Data at different devices and different $V_{BG}$ can be fitted into powerlaw with different exponents $a$. Further studies needs to be done to understand and explain the temperature dependence.

\subsubsection{Device Layout}
\begin{figure*}
\includegraphics{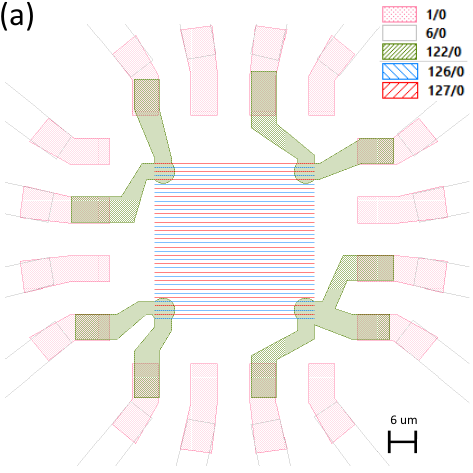}
\centering
\caption{GDSII file used for Device 1. Similar layout was used for Devices 2 and 3. Layer 1: Gold interface electrodes, 2: Bondpads on top of LAO, 122: Virtual electrodes by creating conducting 2DEG on LAO/STO interface by ULV-EBL exposure, 126: Nanowires at 2um spacing, 127: another set of interleaved nanowires that are 2um apart, to make net spacing 1 um. Two more such interleaved exposures were done to make net spacing 250 nm.}
\label{layout}
\end{figure*}

See Fig. \ref{layout}

\subsubsection{ULV-EBL of nanowires}
\begin{figure*}
\includegraphics{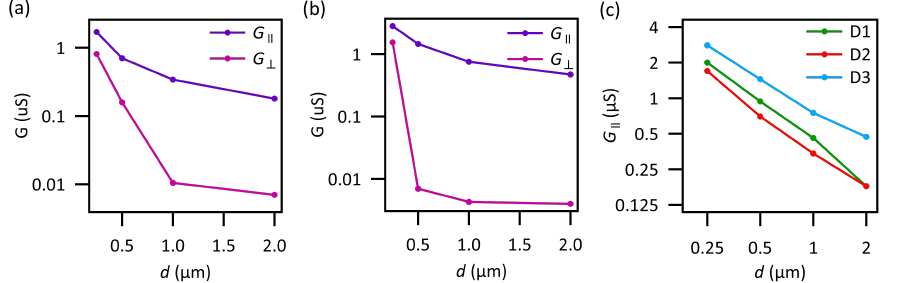}
\centering
\caption{(a) $G_{||}$ and $G_\perp$ as a function of interwire spacing obtained at room temperature during interleaved ULV-EBL exposure of nanowires for (a) Device 2 and (c) Device 3. (c) $G_{||}$ nearly doubles with doubling of interwire spacing as double the number of nanowires make contact to each virtual electrode}
\label{sebl}
\end{figure*}

See Fig. \ref{sebl}

\subsubsection{Back gate tuning of magnetoresistance in device 1}

\begin{figure}
\includegraphics{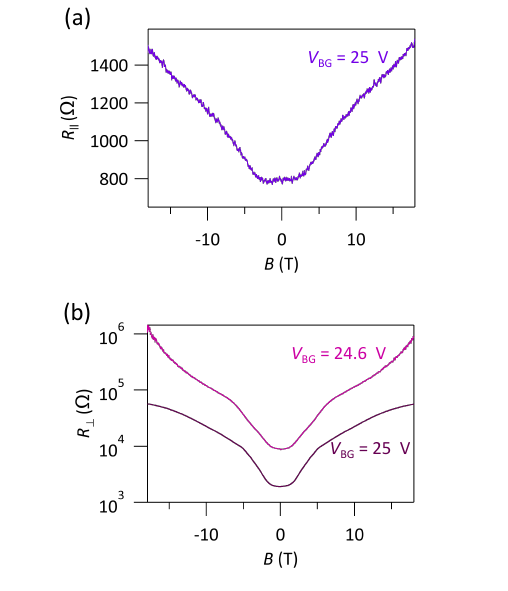}
\centering
\caption{(a) $R_{||}$ and (b)$R_\perp$  as a function of out-of-plane magnetic field for Device 1}
\label{MagnetoRvsVBG}
\end{figure}

\ref{MagnetoRvsVBG}(b) shows that lowering $V_{BG}$ has a drastic effect of increasing the magnetoresistance. This can also be inferred from the fact that $V_0$ increases with on lowering the back gate, rendering the barrier stronger.

\subsubsection{Device 4: Interdigitated device}
\begin{figure*}
\includegraphics{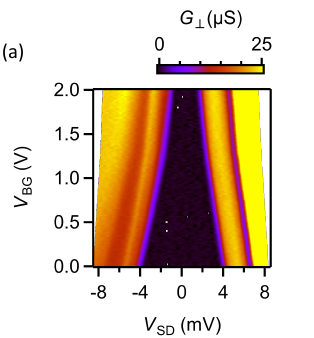}
\centering
\caption{Device 4: (a) $G_\perp$ obtained from IV curves taken at B = 0 T, T $<$ 50 mK as a function of $V_{BG}$}
\label{d4}
\end{figure*}

See Fig. \ref{d4}

\end{document}